\documentclass[pra,aps,reprint,showpacs,superscriptaddress,amsmath]{revtex4-1}
\usepackage{graphicx}
\usepackage{dsfont}

\newcommand{\bra}[1]{\langle #1 | \,}
\newcommand{\ket}[1]{\, | #1 \rangle}

\newcommand{\hlf}{\frac{1}{2}}
\newcommand{\tOm}{\tilde\Omega}
\newcommand{\bOm}{\bar\Omega}
\newcommand{\nth}{\bar{n}_{\mathrm{th}}}

\newcommand{\hrho}{\hat{\rho}}

\begin{document}

\title{Long-range quantum gate via Rydberg states of atoms in
a thermal microwave cavity}

\author{L\H{o}rinc S\'ark\'any}
\author{J\'ozsef Fort\'agh}
\affiliation{CQ Center for Collective Quantum Phenomena and their Applications,
Physikalisches Institut, Eberhard Karls Universit\"at T\"ubingen, 
Auf der Morgenstelle 14, D-72076 T\"ubingen, Germany}

\author{David Petrosyan}
\affiliation{Institute of Electronic Structure and Laser, 
FORTH, GR-71110 Heraklion, Crete, Greece}
\affiliation{Aarhus Institute of Advanced Studies, Aarhus University,
DK-8000 Aarhus C, Denmark}

\date{\today}

\begin{abstract}
We propose an implementation of a universal quantum gate between pairs 
of spatially separated atoms in a microwave cavity at finite temperature. 
The gate results from reversible laser excitation of Rydberg states of 
atoms interacting with each other via exchange of virtual photons through 
a common cavity mode. Quantum interference of different transition paths 
between the two-atom ground and double-excited Rydberg states makes both 
the transition amplitude and resonance largely insensitive to the excitations
in the microwave cavity ``quantum bus'' which can therefore be in any 
superposition or mixture of photon number states. Our scheme for 
attaining ultralong-range interactions and entanglement also applies 
to mesoscopic atomic ensembles in the Rydberg blockade regime and is scalable 
to many ensembles trapped within a centimeter sized microwave resonator.
\end{abstract}


\pacs{
03.67.Lx, 
32.80.Ee, 
37.30.+i, 
}

\maketitle

Quantum interfaces between solid-state devices and cold atoms 
are the backbone of a novel class of hybrid quantum systems 
\cite{Wallquist2009,Bensky2011,Nori2013,Kurizki2015} linking 
fast quantum gates \cite{Schoelkopf2008,Devoret2013} with long-lived 
quantum memories \cite{Lvovsky2009,Fleischhauer2005} and optical 
quantum communication channels \cite{Sergienko2006,Hammerer2010,Sangouard2011}.
Superconducting coplanar waveguide resonators operating in the microwave 
regime have been demonstrated to provide strong coupling between solid-state
superconducting qubits \citep{Blais2004,Sillanpaa2007,DiCarlo2009,Reed2012} 
and to mediate quantum state transfer between superconducting qubits 
and spin ensembles \cite{Zhu2011,Kubo2011,Saito2013}. 
Cold atoms trapped near the surface of an atom chip 
\cite{Treutlein2004,Fortagh2007,HarocheBEC,Bernon2013,Sarkany2014} 
possess good coherence properties and strong optical (Raman) transitions.
Therefore ensembles of trapped atoms or molecules interacting with 
on-chip microwave resonators were proposed as convenient systems 
\cite{Rabl2006,Tordrup2008,Verdu2009,Petrosyan2008,Petrosyan2009} for 
realizing quantum gates and memories as well as optical interfaces. 
A promising approach to achieve strong coupling of atoms to microwave resonators
\cite{Petrosyan2008,Petrosyan2009,Pritchard2014,Hogan2012,Hermann2014,*Teixeira2015} 
is to employ the atomic Rydberg states having huge electric dipole 
moments \cite{Gallagher1994}. 

A common feature of all these schemes is that they require an initially 
empty microwave cavity which should be kept at very low temperatures
of $T \lesssim 100\:$mK. This is routinely achieved with solid-state 
superconducting circuits in dilution refrigerators, but is challenging 
to realize and maintain in combination with ultra-cold atoms \cite{Jessen2013}. 
In turn, atoms are routinely trapped on atom chips at $T \simeq 4\:$K 
\cite{Bernon2013,Hermann2014,*Teixeira2015}, but then the integrated superconducting 
cavities have lower quality ($Q$) factor, and the presence of thermal cavity 
photons and their fluctuations would preclude high-fidelity quantum operations.

Here we present a scalable scheme for cavity-mediated coherent 
interactions between Rydberg states of atoms in a thermal microwave cavity. 
We show that a universal quantum gate between pairs of laser-driven atoms, 
or atomic ensembles in the Rydberg blockade regime 
\cite{Lukin2001,Saffman2010,*Comparat2010}, 
can be achieved with current cold-atom experimental technology 
\cite{Mukai2007,HarocheBEC,Cano2011,Hogan2012,Hermann2014,*Teixeira2015}.

\begin{figure}[t]
\centerline{\includegraphics[width=8.7cm]{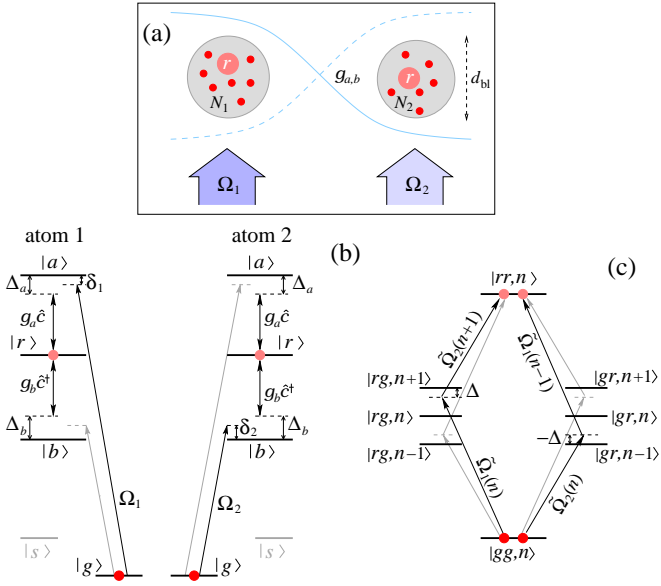}}
\caption{(
Schematics of the system.
(a)~Spatially separated atoms, or superatoms composed of $N_{1,2}$ 
atoms within the Rydberg blockade distance $d_{\mathrm{bl}}$, interact 
with a pair of optical lasers and a common mode of a microwave cavity.
(b)~Level scheme of two atoms interacting with the cavity field 
on the Rydberg transitions $\ket{r} \leftrightarrow \ket{a},\ket{b}$ 
with strengths $g_{a,b}$ and driven by the laser fields 
with Rabi frequencies $\Omega_{1,2}$ on the transitions 
from the ground state $\ket{g}$ to states $\ket{a},\ket{b}$.   
(c)~Under appropriate conditions (see text for details), there 
are two interfering excitation pathways from the two-atom ground state 
$\ket{gg,n}$ to the double-excited Rydberg state $\ket{rr,n}$,
which cancel the dependence of the total transition amplitude 
$\bOm$ on the cavity photon number $n$.}
\label{fig:ALscheme}
\end{figure}

Our work has been inspired by the seminal proposal of S\o{}rensen and M\o{}lmer 
\cite{Sorensen-Molmer1999,*Molmer-Sorensen1999} to realize quantum 
computation and entanglement with ions sharing a common vibrational mode 
subject to thermal fluctuations. Different from the ion trap, in our scheme
the cavity mode exchanging photons with the thermal environment simultaneously
interacts with atoms on the transitions between neighboring Rydberg states, 
playing the role of the ``quantum bus'' for spatially separated qubits
[Fig.~\ref{fig:ALscheme}(a)]. 
Now the photon-number uncertainty affects not only the amplitude of the 
laser-mediated transition between the atomic ground and Rydberg states, 
but also induces fluctuating ac-Stark shifts of atomic levels.
By a suitable choice of the system parameters, we ensure that 
both the multiphoton transition amplitude and its resonant frequency 
are insensitive to the cavity photon number, making the gate operations 
immune to the exchange of photons with the thermal environment, 
which relaxes the need for a low-temperature high-$Q$ cavity. 

We first describe the scheme for individual atoms, and later 
adapt it also to atomic ensembles forming Rydberg ``superatoms'' 
\cite{Lukin2001,Saffman2010,*Comparat2010,Weber2015,*Ebert2015,*Zeiher2015}. 
Consider a pair of identical atoms $1$ and $2$ with the ground state $\ket{g}$
and the highly excited Rydberg state $\ket{r}$ [Fig.~\ref{fig:ALscheme}(b)].
The atoms interact nonresonantly with a common mode of the microwave cavity 
via the dipole-allowed transitions to the adjacent Rydberg states $\ket{a}$
and $\ket{b}$; the corresponding coupling strengths are denoted by $g_{a,b}$.
Two excitation lasers of optical frequencies $\omega_{1,2}$ act on the 
atomic transitions $\ket{g}_1 \to \ket{a}_1$ and $\ket{g}_2 \to \ket{b}_2$
with the Rabi frequencies $\Omega_{1,2}$. For simplicity we assume
for now that each laser interacts only with the corresponding atom 
(see below for the symmetric coupling of both atoms). The total Hamiltonian
of the system in the rotating wave approximation takes 
the form $H = H_c + \sum_{i=1,2} [H_a^{(i)} + V_{ac}^{(i)} + V_{al}^{(i)}]$.
Here $H_c = \hbar \omega_c (\hat{c}^{\dag} \hat{c} + \hlf)$ is 
the Hamiltonian for cavity field with the photon creation $\hat{c}^{\dag}$ 
and annihilation $\hat{c}$ operators in the mode of frequency $\omega_c$, 
$H_a^{(i)} = \hbar \sum_{\mu} \omega_{\mu} \ket{\mu}_i \bra{\mu}$  
is the Hamiltonian of the unperturbed atom $i$ with the Bohr (excitation) 
frequencies $\omega_{\mu}$ of the corresponding energy levels 
$\ket{\mu}$ ($\mu = g,r,a,b$), 
$V_{ac}^{(i)} = \hbar (g_a \hat{c} \ket{a}_i \bra{r} 
+ g_b \hat{c}^{\dag} \ket{b}_i \bra{r} + \mathrm{H. c.})$ describes 
the atom-cavity interactions, and
$V_{al}^{(1)} = \hbar \Omega_1 e^{-i \omega_1 t} \ket{a}_1 \bra{g} + \mathrm{H. c.}$
and 
$V_{al}^{(2)} = \hbar \Omega_2 e^{-i \omega_2 t} \ket{b}_2 \bra{g} + \mathrm{H. c.}$
describe the laser driving of atoms 1 and 2, respectively. 

We assume that the two atoms initially in the ground state $\ket{g}$ are in 
spatially separated traps at equivalent positions, close to field antinodes
of the microwave cavity containing an arbitrary number of photons $n$.
Our aim is to achieve coherent oscillations between the collective states
$\ket{gg,n}$ and $\ket{rr,n}$ with maximal amplitude and oscillation
frequency which do not depend on $n$. Denoting the detunings 
of the laser fields by $\delta_1 = \omega_{ag} - \omega_1$ 
and $\delta_2 = \omega_{bg} - \omega_2$ and taking Rabi frequencies 
$\Omega_{1,2} \ll |\delta_{1,2}|$, we adiabatically eliminate the
intermediate atomic states $\ket{a}_1$ and $\ket{b}_2$, 
obtaining two-photon Rabi frequencies 
$\tOm_1(n) = \frac{\Omega_1 g_a \sqrt{n+1}}{\delta_1}$ and 
$\tOm_2(n) = \frac{\Omega_2 g_b \sqrt{n}}{\delta_2}$ on 
the transitions $\ket{gg,n} \to \ket{rg,n+1}$ and 
$\ket{gg,n} \to \ket{gr,n-1}$ accompanied by addition and 
subtraction of a cavity photon, respectively [Fig.~\ref{fig:ALscheme}(c)].    

We next take large and unequal detunings $\Delta_a = \omega_{ar} - \omega_c$ 
and $\Delta_b = \omega_{rb} - \omega_c$ of the cavity field from the transition
resonances between the atomic Rydberg states, $|\Delta_{a,b}| \gg g_{a,b}$.
To avoid cavity-mediated F\"orster resonances 
$\ket{rr,n} \to \ket{ab,n},\ket{ba,n}$ \cite{Petrosyan2008}, we require
that $|\Delta_a - \Delta_b| \gg g_a g_b \left| \frac{n+1}{\Delta_b} 
- \frac{n}{\Delta_a}\right|$ for all $n \lesssim n_{\mathrm{max}}$,
where $n_{\mathrm{max}}$ is the maximal possible cavity photon number
[typically $n_{\mathrm{max}} \approx 2 \nth$ for a thermal field with the 
mean number of photons $\nth = (e^{\hbar \omega_c/k_{\mathrm{B}} T} -1)^{-1}$]. 
If we now choose the two-photon detunings 
$\Delta_{1,2} \approx \delta_{1,2} \mp \Delta_{a,b}$ of states 
$\ket{rg,n+1}$ and $\ket{gr,n-1}$ to have equal magnitude 
but opposite sign, $\Delta_{1} = -\Delta_{2} = \Delta \gg \tOm_{1,2}$,
we can also eliminate these intermediate states and obtain 
resonant multiphoton transitions between states $\ket{gg,n}$ 
and $\ket{rr,n}$ involving two laser photons and an exchange of a (virtual)
cavity photon between the two atoms. With two equivalent excitation paths
from $\ket{gg,n}$ to $\ket{rr,n}$ [Fig.~\ref{fig:ALscheme}(c)], 
the resulting transition amplitude (effective Rabi frequency) is
\begin{equation}
\bOm = \frac{\tOm_1(n) \tOm_2(n+1)}{\Delta_1(n)}
+  \frac{\tOm_2(n) \tOm_1(n-1)}{\Delta_2(n)}  
= \frac{\Omega_1 \Omega_2 g_a g_b}{\delta_1 \delta_2 \Delta} . \label{Omega4eff} 
\end{equation}
This is the photonic cavity analog of the S\o{}rensen-M\o{}lmer scheme
\cite{Sorensen-Molmer1999,*Molmer-Sorensen1999} for the ion trap with phonons.
The critical question now is how to precisely tune the detunings $\Delta_{1,2}$ 
and achieve the multiphoton resonance $\ket{gg,n} \leftrightarrow \ket{rr,n}$ 
for any $n$. 

From the perturbative analysis, we obtain that the detunings 
$\Delta_{1,2}(n)$ depend on the cavity photon number $n$ through 
the second order (ac Stark) shifts of levels $\ket{r}_{1,2}$, 
\begin{eqnarray*}
\Delta_{1}(n) &\simeq & \delta_{1} + \frac{\Omega_{1}^2}{\delta_{1}} 
- \Delta_{a} - \frac{g_a^2(n+1)}{\delta_{1}} + \frac{g_b^2(n+2)}{\Delta_b}, \\
\Delta_{2}(n) &\simeq & \delta_{2} + \frac{\Omega_{2}^2}{\delta_{2}} 
+ \Delta_{b} - \frac{g_a^2(n-1)}{\Delta_a} - \frac{g_b^2n}{\delta_{2}} .
\end{eqnarray*}
With an appropriate choice of $\delta_{1,2}$ and $\Omega_{1,2}$, 
and requiring that $\frac{g_a^2}{\Delta_a} = \frac{g_b^2}{\Delta_b}$,
the $n$-dependence of the detunings is greatly suppressed, 
$\Delta_{1,2}(n) \simeq \Delta_{1,2}(0) 
\left(1+\frac{g_{a,b}^2}{\Delta_{a,b}^2} n \right)$, and we can satisfy 
the condition $|\Delta_{1}(n) + \Delta_{2}(n)| \ll \bOm$ for any $n$. 
This leads to $\bOm$ that only weakly depends on $n$, 
$\bOm (n) \simeq \bOm(0) 
\left( 1 - \frac{g_a^2}{\Delta_a^2} n \right)$.
In order to ensure the multiphoton resonance on the transition 
$\ket{gg,n} \to \ket{rr,n}$, we also need to consider higher-order 
level-shifts of $\ket{gg,n}$ and $\ket{rr,n}$. 
To fourth order in the laser and cavity field couplings, 
the largest contribution to the level shift of $\ket{gg,n}$ is given 
by $S_{gg}(n) = \frac{\Omega_1^2 g_a^2 (n+1)}{\delta_1^2 \Delta_1(n)}
+ \frac{\Omega_2^2 g_b^2 n}{\delta_2^2 \Delta_2(n)}$, which, remarkably,
has the same structure as $\bOm$. Since $\Delta_{1}(n) \simeq - \Delta_{2}(n)$,
we can choose $\frac{\Omega_1 g_a}{|\delta_1|}=\frac{\Omega_2 g_b}{|\delta_2|}$ 
to make $S_{gg}$ nearly independent on $n$ and absorb it into $\Delta_{1,2}(n)$.
Finally, the fourth-order shifts of $\ket{rr,n}$, 
$S_{rr}(n) \propto \frac{g_{a,b}^4 n^2}{\Delta_{a,b}^3}$ \cite{Petrosyan2008}  
are small in comparison and can therefore be neglected, which we verify 
below through exact numerical simulations for the complete system. 

In general, the atom-cavity couplings $g_a$ and $g_b$ are not equal, 
since they are proportional to the electric dipole matrix elements 
on different transitions $\ket{r} \to \ket{a}$ and $\ket{r} \to \ket{b}$,
while the corresponding detunings $\Delta_a$ and $\Delta_b$ can be tuned 
by static electric (Stark) or magnetic (Zeeman) fields \cite{Gallagher1994}. 
These, together with the flexibility to choose the laser detunings 
$\delta_{1,2}$ and Rabi frequencies $\Omega_{1,2}$, allows us to satisfy all of 
the above conditions for $n$-independent resonant Rabi oscillations between 
states $\ket{gg}$ and $\ket{rr}$. We can estimate the maximal attainable 
oscillation frequency $\bOm$, assuming that the main limiting factor is the 
atom-cavity coupling strengths $g_{a,b}$ \cite{Petrosyan2009,Hogan2012}, since 
the laser Rabi frequencies can be collectively enhanced in the superatom regime
\cite{Lukin2001,Saffman2010,*Comparat2010,Weber2015,*Ebert2015,*Zeiher2015}.
Recall that we require the intermediate state detunings 
$|\Delta_{1,2}(n)| \gg \tOm_{1,2}(n)$ for each $n \lesssim n_{\mathrm{max}}$. 
Then, with $\eta = |\delta_2|/\Omega_2$ and $\tilde\eta = |\Delta_1|/\tOm_1$
and all of the above conditions satisfied, we obtain that
$\bOm_{\text{max}} \leq \frac{g_b}{\eta\bar \eta\sqrt{n_{\text{max}}+1}}$. 
As an estimate, assuming $g_{a,b} \simeq 2\pi \times 10$ MHz, 
$\eta,\tilde\eta \simeq 10$ and a cavity mode with 
$\omega_c \simeq 2\pi \times 15\:$GHz at $T=4\:$K, leading to
$\nth \approx 5$ ($n_{\mathrm{max}} \approx 2 \nth$),
we have $\tOm_{\text{max}} \approx 2\pi \times 30\:$kHz. 

Note that, similarly to the ion trap scheme 
\cite{Sorensen-Molmer1999,*Molmer-Sorensen1999}, our scheme would work 
in exactly the same way for the symmetric coupling of both atoms 
to both excitation lasers [Fig.~\ref{fig:ALscheme}(b)]. This opens 
two new excitation paths from $\ket{gg,n}$ to $\ket{rr,n}$ via states 
$\ket{rg,n-1}$ and $\ket{gr,n+1}$ [Fig.~\ref{fig:ALscheme}(c)], which
enhances the transition amplitude $\bOm$ by a factor of two. Care should 
be taken only to properly account for, and compensate, the additional 
second- and fourth-order levels shifts of the atomic ground states.

\begin{figure}[t]
\centerline{\includegraphics[width=8cm]{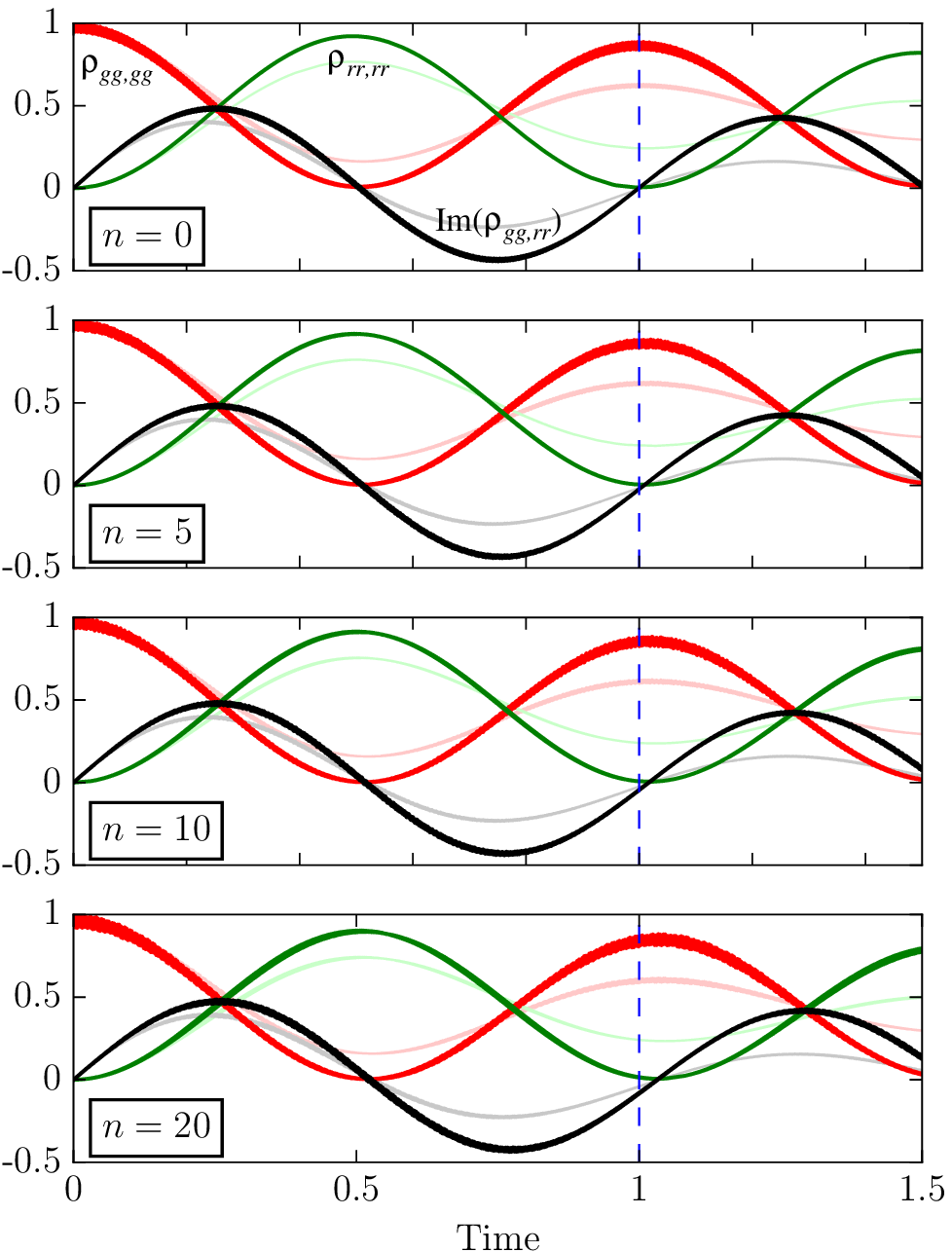}}
\caption{(
Rabi oscillation between states $\ket{gg,n}$ 
and $\ket{rr,n}$ for the cavity photon numbers $n=0,5,10,20$. 
Thick (full) lines show coherent oscillations of populations 
$\rho_{gg,gg} = \bra{gg} \hrho \ket{gg}$, 
$\rho_{rr,rr} = \bra{rr} \hrho \ket{rr}$ and 
coherence $\rho_{gg,rr} = \bra{gg} \hrho \ket{rr}$ 
for the Rydberg state decay $\Gamma=0.142 \,[\bOm(0)/2\pi]$
and no dephasing, $\gamma=0$, 
while thinner (shaded) lines show damped oscillations 
of the same quantities for large dephasing $\gamma=0.4 \, [\bOm(0)/2\pi]$. 
Parameters are $\Omega_1 = 56.50, \Omega_2 = 60.00, \delta_1 = 663.8, 
\delta_2 =  -742.0, g_a= 9.5, g_b= 10.0, \Delta_a = 722.0, \Delta_b=800.0$
($\times 2\pi\:$MHz), leading to $\bOm(0) \approx 2 \pi \times 21.1\:$kHz. 
Time is in units of $[2\pi/\bOm(0)]$.}
\label{fig:Rabi-osc}
\end{figure}

To validate our perturbative calculations, we numerically solve 
the master equation $\frac{\partial}{\partial t} \hrho 
= -\frac{i}{\hbar} [H, \hrho]$ for the density operator $\hrho$
using the exact Hamiltonian $H$ for the pair of atoms initially 
in the ground state $\ket{g}$ and the cavity field with $n$ photons. 
Results for different $n$ are shown in Fig.~\ref{fig:Rabi-osc}, 
which verifies that with a proper choice of parameters, 
the transition resonance $\ket{gg,n} \leftrightarrow \ket{rr,n}$ 
and the effective Rabi frequency $\bOm$ can simultaneously be made 
nearly-independent on the number of photons in the cavity. We have 
examined various Fock, coherent and thermal states as the initial 
states of the cavity field, all yielding very similar results. 

We include the realistic relaxation processes affecting the Rydberg states 
of atoms using the standard Liouvillians \cite{PLDP2007} with the Lindblad 
generators $\hat{L}^{(i)}_{\nu} = \sqrt{\Gamma}\ket{g}_i\bra{\nu}$ for the 
decay with rate $\Gamma$ (assumed the same for all $\nu = r,a,b$), 
and $\hat{L}^{(i)}_{g} = \sqrt{\gamma/2} (\ket{g}_i\bra{g} 
- \sum_{\nu} \ket{\nu}_i\bra{\nu})$ for the dephasing 
(with respect to the ground state) with rate $\gamma$. 
In Fig.~\ref{fig:Rabi-osc} we show strongly damped Rabi oscillations caused 
by the single-atom decay $\Gamma$ and relatively large dephasing $\gamma$ 
comparable to $\bOm$. 

\begin{figure}[t]
\centerline{\includegraphics[width=7cm]{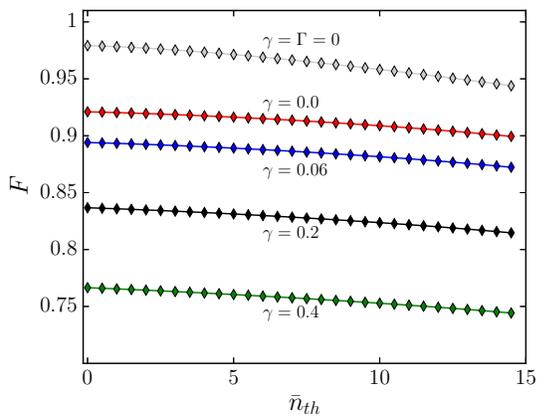}}
\caption{
Transfer fidelity $F$ at time $t_{\pi} = \pi/\bOm(0)$ 
versus the mean number $\nth$ of thermal photons in the cavity, for the
dephasing rates $\gamma = 0,0.06,0.2,0.4$ and the decay rates 
$\Gamma=0$ (uppermost reference curve) and 
$\Gamma=0.142$ (all other curves), in units of $[\bOm(0)/2\pi]$.
Other parameters are as in Fig.~\ref{fig:Rabi-osc}.}
\label{fig:fidelity}
\end{figure}

In Fig.~\ref{fig:fidelity} we plot the fidelity 
$F = \sum_n p(n|\nth) \bra{rr,n} \hrho(t_{\pi}) \ket{rr,n}$ 
of transfer $\ket{gg} \to \ket{rr}$ at time $t_{\pi} = \pi/\bOm(0)$
(effective $\pi$-pulse for $n=0$) as a function of the dephasing rate 
$\gamma$ and the mean number of thermal photons $\nth$ which determines 
the photon number probability distribution $p(n|\nth)=\nth^n/(\nth+1)^{n+1}$.
We observe that large dephasing detrimentally affects the transfer 
fidelity by damping the amplitude of Rabi osculations between 
$\ket{gg}$ and $\ket{rr}$. On the other hand, the fidelity only weakly 
and nearly linearly decreases with increasing the cavity photon number, 
due to the slight decrease of the effective Rabi frequency $\bOm(n)$ 
discussed above. This can be compensated by appropriately correcting 
the transfer time $t_{\pi}$ if $\nth$ is approximately known. 

We now discuss the implications of our results for quantum information 
applications. Each atom can encode a qubit as a coherent superposition 
of long-lived states $\ket{g}$ and $\ket{s}$ [Fig.~\ref{fig:ALscheme}(b)]. 
Gate operation can be performed on any pair of atoms within the cavity 
by addressing the desired atoms with focused laser pulses $\Omega_{1,2}$.
Assuming that state $\ket{s}$ is decoupled from the laser field(s),
the two atom state $\ket{ss}$ is immune to the lasers.
If only one of the atoms is initially in 
state $\ket{g}$ and the other atom is in $\ket{s}$, both atoms remain 
in their initial states due to the absence of multiphoton resonances 
to any Rydberg state. Finally, if both atoms are in state $\ket{g}$, 
the application of lasers for time $t_{2\pi} = 2\pi/\bOm$ will drive 
a complete Rabi cycle on the transition $\ket{gg} \leftrightarrow \ket{rr}$,
resulting in the sign change of $\ket{gg}$. Since the other initial states 
remain unaltered, this transformation corresponds to the universal two-qubit 
\textsc{cphase} logic gate \cite{PLDP2007,Nielsen2000}.  

Our scheme is also applicable to ensembles of trapped atoms in the 
Rydberg blockade regime \cite{Lukin2001,Saffman2010,*Comparat2010}. 
Individual ensembles, each containing $N_i \gg 1$ atoms, can encode qubits 
in the collective ground $\ket{G} \equiv \ket{g_1,g_2,\ldots,g_{N_i}}$ 
and symmetric single spin-flip (hyperfine) 
$\ket{S} \equiv \frac{1}{\sqrt{N_i}}\sum_{k=1}^{N_i} 
\ket{g_1,\ldots,s_k,\ldots, g_{N_i}}$ states. For optically dense ensembles, 
the qubit encoding superposition of states $\ket{G}$ and $\ket{S}$
can be reversibly mapped onto photonic qubits via stimulated
Raman techniques \cite{Fleischhauer2005}. We assume that each 
ensemble of linear dimension smaller than a certain blockade distance 
$d_{\mathrm{bl}}$ [Fig.~\ref{fig:ALscheme}(a)] behaves as a Rydberg superatom 
\cite{Lukin2001,Saffman2010,*Comparat2010,Weber2015,*Ebert2015,*Zeiher2015}
wherein multiple excitations are suppressed by the strong Rydberg-state
interactions. This allows implementation of arbitrary single-qubit 
and universal two-qubit quantum gates as follows. In each ensemble, 
before and after the gate execution, we apply a single-atom $\pi$-pulse 
on the transition $\ket{s} \to \ket{r'}$, where $\ket{r'}$ is a Rydberg 
state which can block excitation of any other atom from state $\ket{g}$ 
to states $\ket{r'},\ket{r}$ (and, possibly, to $\ket{a},\ket{b}$)
due to the strong dipole-dipole or van der Waals interactions 
\cite{Saffman2010,*Comparat2010}. This operation reversibly maps 
the qubit state $\ket{S}$ onto the symmetric single Rydberg excitation 
state $\ket{R'} \equiv \frac{1}{\sqrt{N_i}}\sum_{k=1}^{N_i} 
\ket{g_1,\ldots,r'_k,\ldots, g_{N_i}}$. Single qubit transformations are 
then performed in the two-state subspace of $\ket{G}$ and $\ket{R'}$
by resonant lasers with collectively enhanced Rabi frequencies 
$\Omega = \sqrt{N_i} \Omega^{(1)}$, where $\Omega^{(1)}$ is the 
single-atom Rabi frequency on the transition $\ket{g} \leftrightarrow \ket{r'}$ 
\cite{Lukin2001}. 
For the two qubit operations, any pair of atomic ensembles $i,j$ 
trapped within the cavity can be addressed by appropriate lasers
with collectively enhanced Rabi frequencies 
$\Omega_{1,2} = \sqrt{N_{i,j}} \Omega_{1,2}^{(1)}$. Then, during time $t_{2\pi}$, 
the initial state $\ket{G}_i\ket{G}_j$ will undergo a complete Rabi cycle 
to $\ket{R}_i\ket{R}_j$ and back, acquiring a $\pi$ phase shift 
(sign change), assuming that in each ensemble multiple excitations of 
$\ket{r}$ are blocked by strong Rydberg-state interactions. If any, 
or both, of the ensembles were initially in state $\ket{S}$ mapped 
onto $\ket{R'}$, the atom in $\ket{r'}$ would preclude the transfer 
of ground state atoms $\ket{g}$ to $\ket{r}$, and therefore such initial 
states will remain unaltered. This completes the implementation of the 
\textsc{cphase} logic gate with Rydberg superatoms. 

To conclude, our scheme to implement long-range quantum gates 
is feasible with present-day experimental technologies 
\cite{Bernon2013,Hogan2012,Jessen2013,HarocheBEC,Hermann2014,*Teixeira2015}
involving optical excitation of Rydberg states of trapped atoms 
and their interactions with microwave resonators. This gate is largely 
insensitive to the number of cavity photons and it can therefore operate 
in a finite-temperature microwave cavity with modest photon lifetimes 
$1/\kappa \gtrsim 1 \:\mu$s. 
The main decoherence mechanisms reducing the achievable gate 
fidelity $F \simeq 1- (\Gamma+\gamma) \tau$ during time 
$\tau = 2\pi/\bOm(0) \sim 50\:\mu$s are the decay $\Gamma \simeq 3\:$kHz 
of Rydberg states and dephasing $\gamma \lesssim 5\:$kHz of the non-resonant 
multiphoton transitions [with $\Delta \gg \tOm_{1,2}(n)$], which are slow 
by construction \cite{Sorensen-Molmer1999,*Molmer-Sorensen1999}.
It would thus be interesting to explore the near-resonant excitations
[with $\Delta/\tOm_{1}(0) = 2 \sqrt{m}$ ($m=1,2,\ldots$)] analogous to 
the fast-gate regime of the ion traps \cite{Sorensen-Molmer2000,*Sackett2000}.
Unlike our present gate, however, such a fast-gate scheme will be sensitive 
to the change of cavity photon number \cite{Roos2008} during the gate time 
$\tau = 2\pi \sqrt{m}/\tOm_{1}(0) \sim 10\:\mu$s requiring cavities 
with longer photon lifetimes $1/\kappa \gg \tau$.

\begin{acknowledgments}
We thank Klaus M\o{}lmer for useful comments and suggestions.
This work was supported by the Deutsche Forschungsgemeinschaft SFB TRR21,
the European Union FP7 STREP project HAIRS and H2020 FET Proactive project RySQ.
D.P. is grateful to the University of Kaiserslautern for hospitality 
and to the Alexander von Humboldt Foundation for support 
during his stay in Germany. 
\end{acknowledgments}

\bibliography{refs4cavTrr}

\end{document}